# Multiscale transform based seismic reflectivity inversion using convolutional neural network

John Castagna, *University of Houston*

Oleg Portniaguine, Gabriel Gil, Arnold Oyem, Chen Liang, *Lumina Geophysical, Houston, U.S.A.*

Corresponding author: Chen Liang, E-mail: chen.liang@luminageo.com

**Abstract**

The Multiscale Fourier Transform of a seismic trace performs time-frequency analyses over a range of window lengths. The variation in window length captures local and global relative amplitudes between events, thereby allowing reflectivity inversion that is independent of the amplitude spectrum of the seismic wavelet. As the temporal and spatial variation of the actual seismic wavelet in seismic reflection data is poorly known, this approach has many advantages over conventional seismic reflectivity inversion. No wavelet extraction is performed. Thus, the inversion for reflectivity can be conducted without well control, seismic ties, or time-depth functions. The inversion is sparse, so no starting model is needed. Furthermore, as no wavelet is required, the inversion can be applied directly to depth migrated data. The phase of the wavelet is constrained by the assumption of sparse reflectivity and thus works best when earth impedance structure is blocky. Trace integration of the inverted reflectivity provides bandlimited impedance which compares very favorably to well-log bandlimited impedance for both synthetic and real data cases.

**Introduction**

Conventional seismic inversion requires the seismic wavelet to be known *a priori*. Determining this wavelet can be a slow and tedious process subject to serious error both at wells with imperfect impedance and time-depth information and between wells where the earth impedance structure and spatial wavelet variation is unknown. Statistical wavelet extraction may rely on invalid assumptions such as white reflectivity and zero phase. Various reflectivity inversion schemes (e.g., Zhang and Castagna, 2011) pursue improved layer resolution but require reliably estimated seismic wavelets to establish inversion kernels. In training neural networks to predict reflectivity from seismic traces, we realized that the long-held belief that the seismic wavelet and its variation in time and space is required to invert seismic reflection data is not true if the inversion problem is structured properly. Liang et al. (2017) demonstrates lost frequencies in seismic reflectivity can be restored with physical principles, aligning with the core idea of using artificial intelligence to uncover and extract features that are associated with fundamental physics. In this paper, we explain how we came upon that realization, and we show examples of inversion without the wavelet on synthetic data to prove the concept and on real data to show its practicality.

**Theory and Method**

Locci et al (2018) introduced the Multiscale Fourier Transform which is a generalization of the Wavelet Transform. The unscaled multiscale transform is simply a Short-Time Fourier Transform over variable window lengths:

$$S(\tau, f, \sigma) = \int_{-\infty}^{\infty} s(t) \, w\left(\frac{t-\tau}{\sigma}\right) e^{-i2\pi f(t-\tau)} \, dt \qquad (1)$$

where, $t$ is seismic record time, $f$ is frequency, $s(t)$ is the seismic trace, $w(t)$ is the mother window, $\tau$ is the window position as it slides along the trace, $\sigma$ is the window length, and $S(\tau, f, \sigma)$ is the time-frequency-scale analysis that maps a single amplitude versus time trace into a 3D dimensional function of amplitude versus window position, frequency, and window length. Note that the complex sinusoidal basis functions, $e^{-i2\pi ft}$, move with the window position. Figure 1 illustrates the Multiscale Transform with time-frequency analyses at a few window lengths.

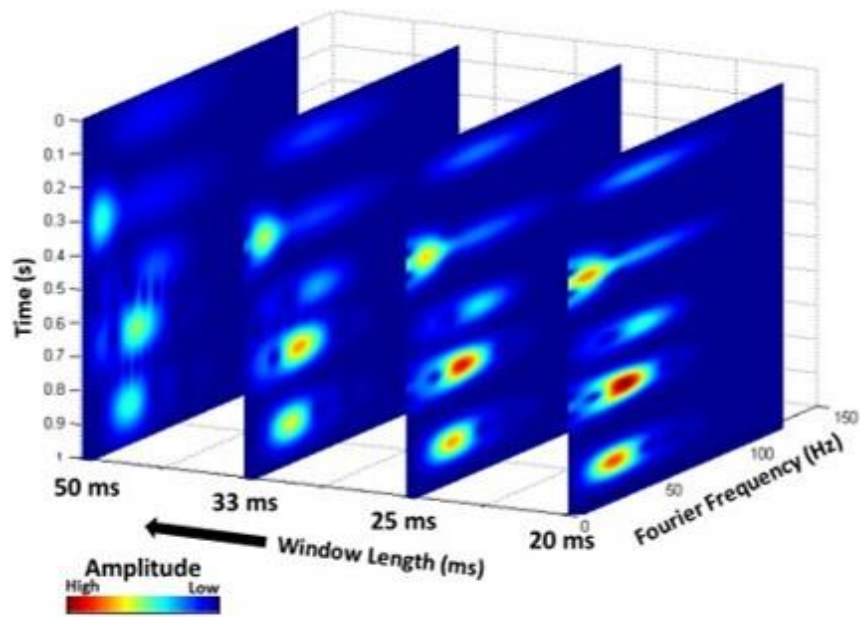

*Figure 1:* *Multiscale Fourier Transform of a synthetic seismic trace using boxcar windows of various lengths. Displayed time-frequency analyses are amplitude spectra.*

The Multiscale Transform reduces to a Continuous Wavelet Transform (e.g., Chakraborty and Okaya; 1995) when the scale, $\sigma$, is forced to be a function of frequency, $f$. The windows at the frequency dependent scales times the sinusoidal basis functions constitute the complex wavelet dictionary. In our implementation here, we allow all achievable scales given the trace length at each frequency.

To invert the seismic trace, we take ratios of the spectral amplitudes at various scales and refer to these as features, $m_i$, of the Multiscale Transform. These are directly inverted using a forward model and a sparsity assumption (see for example Zhang and Castagna, 2011) by minimizing the objective function:

$$\|m_i(\tau, f, \sigma) - m_i'(\tau, f, \sigma)\|_2 + \lambda \|r(t)\|_1 \qquad i=1,n! \qquad (2)$$

where $r(t)$ is the inverted reflectivity series and $\lambda$ is a regularization parameter. The subscripts 1 and 2 refer to the L1 and L2 norms respectively. The $m_i'$ are the modeled $m_i$ which are functions of the observed ratios of spectral amplitudes (trigonometrically transformed to avoid instability caused by division by zero) at each frequency for $n$ window lengths centered at each window position. Not all ratios are useful, especially in the presence of a time or space varying wavelet. A more parsimonious set of ratios that locally largely cancel the wavelet amplitude spectrum can be found by training a neural network to predict reflectivity on synthetic data with various wavelets with different amplitude spectra (see figure 2). Useful $m_i$ can be found by inspection of the aggregated neural network weights. Once these spectral features are identified, no additional training is required, and the problem becomes a classical geophysical inversion determining the sparse reflectivity series that matches those features via a forward model.

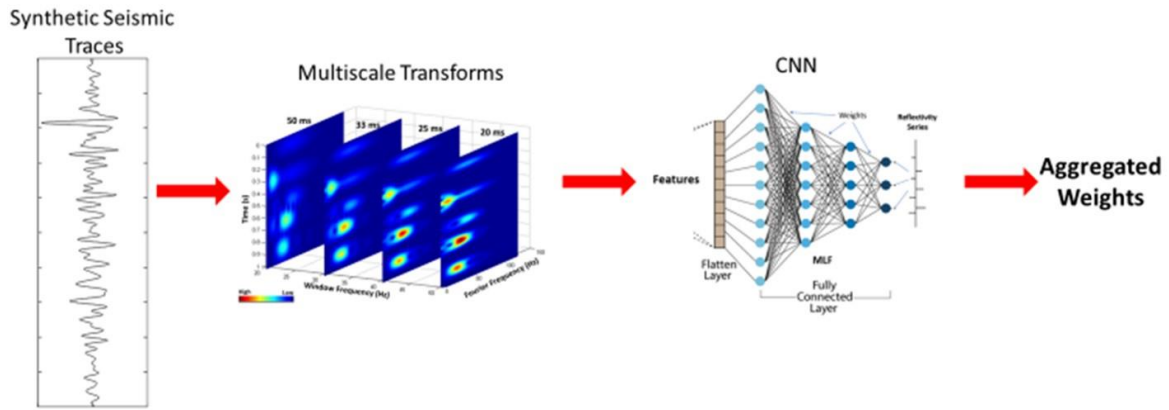

*Figure 2:* *Schematic illustrating the identification of wavelet independent features of the multiscale transform by training a multilayer feedforward (MLF) convolutional neural network (CNN) to predict reflectivity series using synthetic seismic traces with varying wavelets and known reflectivity. The convolutionally pooled features input to the MLF neural network are combinations of the synthetically modeled $m_i$. As the training set is expanded by varying the wavelets and modeled reflectivity series, many $m_i$ drop out. The most useful elements are identified by inspecting the aggregated weights. The CNN architecture diagram is modified from a blog by N. Shariar* https://nafizshahriar.medium.com/what-is-convolutional-neural-network-cnn-deep-learning-b3921bdd82d5

Any forward modeling method can be used in the training and inversion. The synthetic data can be post-stack or pre-stack and can be 3D in time or depth (where record times and time shifts can be replaced by depths). In the implementation shown here, we utilize a 1D primaries-only convolutional model in time which assumes that processing has brought the data to a point where it can be simulated well with this oversimplified model. Inadequacies in the forward model, well logs, or seismic processing can presumably be identified after the fact at validation wells as no well information is used in the inversion (see below). Notably, once the useful features are identified, classical geophysical inversion is applied to those features, and no retraining is required for any specific instance. That such inversion is theoretically possible is demonstrated by Liu et al. (2022).

A first pass of reflectivity inversion is applied assuming the seismic data are zero phase. This results in a reflectivity series whose amplitude spectrum time-frequency-scale analysis is correct, but that may have incorrect phase. By "wavelet phase" we mean a group phase that is approximately constant in the frequency domain over the band of the data but may slowly change vertically in the time domain and laterally, along with the changing wavelet amplitude spectrum. Such wavelets can be obtained in the time domain as a combination of real and imaginary parts of complex wavelets as in Portniaguine and Castagna (2004). Again, assuming a convolutional model, the time and space varying amplitude spectrum of the wavelet is that whose local product with the reflectivity spectrum best matches the seismic data signal spectrum plus uncorrelated noise. Then, given the local amplitude spectrum of the wavelet, the wavelet phase is optimized, varying the wavelet phase rotation and reinverting using a minimum support inversion (Portniaguine and Castagna, 2004; Portniaguine and Castagna, 2005; Chopra et al, 2006). Previously, given the wavelet amplitude spectrum, van der Baan (2008) maximized kurtosis to determine the wavelet phase. Similarly, in our case, the phase value that locally yields the sparsest reflectivity is selected. The correctness of the wavelet phase obtained in this fashion depends on two major assumptions (1) at any position, the wavelet phase is approximately constant over the dominant band of the data in the frequency domain, and (2) the earth locally has a blocky impedance structure. Violation of these assumptions can produce instability in the extracted wavelet phase. Relying on an additional assumption (3) that the wavelet phase is slowly varying in the time domain, such instabilities can be combatted using physical constraints. Once phase is estimated in this way, a final inversion pass with a relaxed sparsity constraint is applied to best match the observed seismic trace.

Below we show tests of the method on synthetic and real data. Note that no additional training or parameterization is utilized for these tests.

**Synthetic Tests**

For our first test, we compute reflectivity from a well log, and produce synthetics using bandpass Ormsby wavelets with a 5 Hz low pass, varying high pass frequencies, and with one octave ramps on the low and high ends. We then integrate the well log and inverted reflection coefficients to produce inverted bandlimited impedance. Figure 3 compares inverted bandlimited impedance with the seismic wavelet (top row) and with perfectly known wavelets (bottom row). Notably, the bandlimited impedance inversions with exactly known *a priori* wavelets appear to be no better than the inversions with unknown wavelets when the data has broad bandwidth. Of course, in practice, the wavelets are rarely exactly known and resulting error in the use of a supposed known wavelet are ignored in this comparison. As more high frequencies are included in the synthetics, the inversions show improved detail and resolution. For the case of the broadest band 45 Hz high pass wavelet, the inversions with or without a known wavelet are in close accord with the well log (see Figure 4).

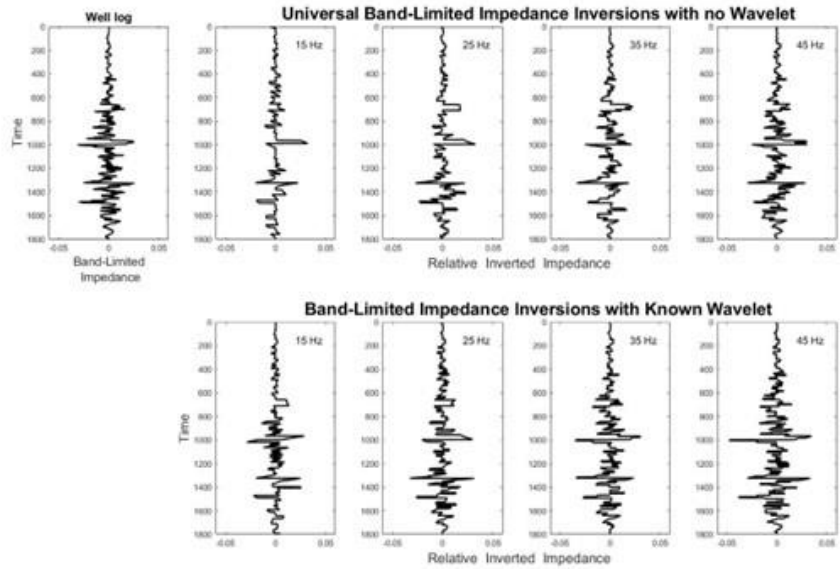

*Figure 3:* Synthetic test #1. Well log bandlimited impedance, and inverted bandlimited impedances for synthetic data using various zero-phase Ormsby bandpass wavelets with high pass frequencies varying from 15 Hz to 45 Hz. The top row shows inversions without the wavelet while the bottom row shows sparse inversions with perfect knowledge of the wavelet. Especially for the broadest band and highest frequency wavelet (45 Hz) the inversion with a known wavelet is no better than the inversion without a wavelet which compares well to the well log bandlimited impedance.

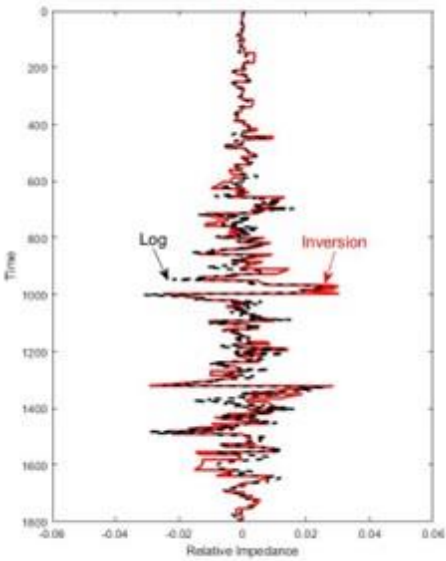

*Figure 4: Synthetic test #1 with zero-phase bandpass wavelet and 45 Hz high pass. Inverted bandlimited impedance without the wavelet (red curve) overlain with well log (black dashed curve).*

Synthetic test #2 shown in Figure 5 is like the first test, but with an unknown 60° Ormsby wavelet in the synthetic that is unknown to the inversion. The results are comparable to the zero-phase wavelet result but slightly less stable, possibly due to violation of the sparsity assumption in the phase determination.

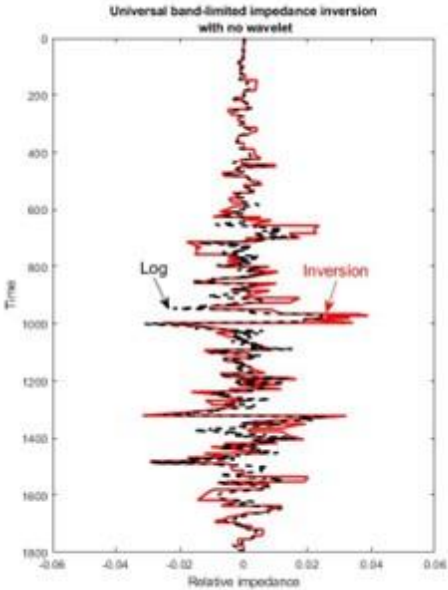

*Figure 5:* Overlay of bandlimited well log and inverted impedance. The synthetic has a 60° Ormsby wavelet with a 45 Hz high pass. The inversion for the most part handles the unknown phase well.

**Real Data Test**

Figure 6 shows a seismic line and its reflectivity inversion without a known or extracted wavelet. No well information or other starting model is used in the inversion. The reflectivity inversion shows greater apparent geological detail and improved layer resolution. Apparent time resolution improves from about 20 ms on the original data to about 10 ms on the reflectivity inversion. Similar reflectivity inversions presented by Puryear and Castagna (2008) and Zhang and Castagna (2011) required that the seismic wavelet and its spatial and temporal variation be known *a priori*.

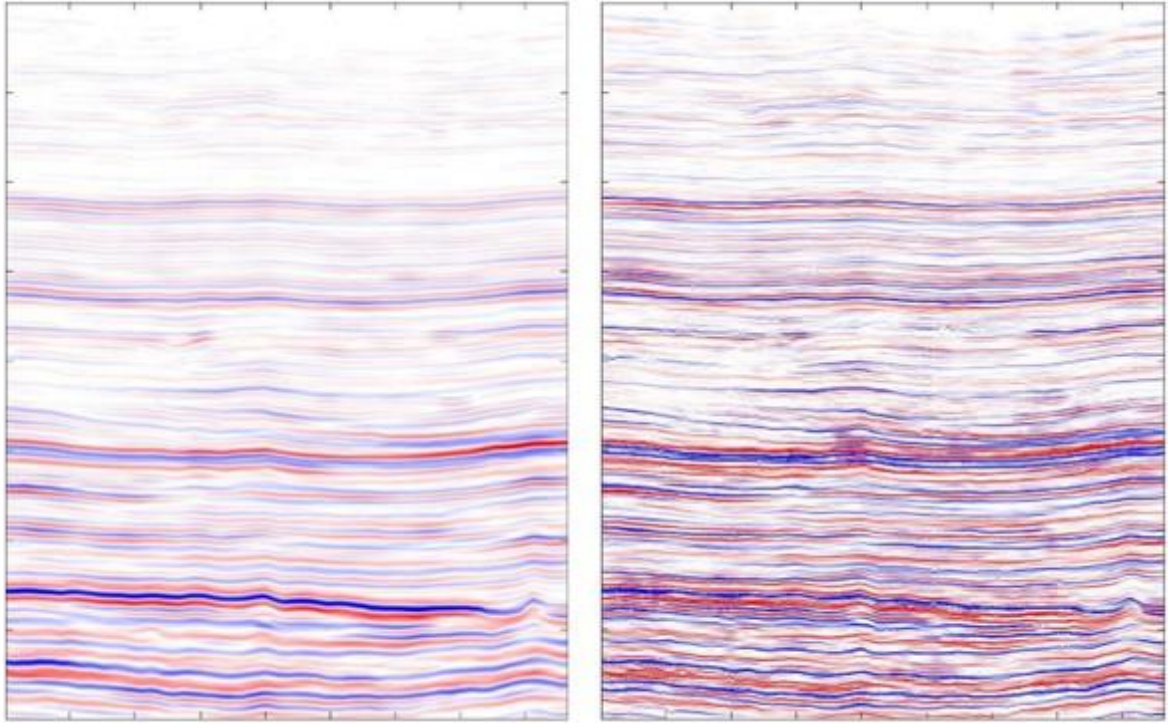

*Figure 6: Left – Original Seismic Section. Right – Reflectivity inversion without an a priori wavelet. Greater geological detail and resolution is apparent on the reflectivity inversion. Seismic times and spatial positions are masked. The timing lines are 200 ms. On the seismic line, red is soft and blue is hard. On the reflectivity inversion, red is a negative reflection coefficient and blue is a positive reflection coefficient.*

Figure 7 shows close ups of the reflectivity inversion compared to impedance logs at wells. Interfaces are precisely recognized and pinching out of layers between wells is verified by the logs. Impedance transitions evident on the logs are indicated by series of same signed reflection coefficients on the reflectivity inversion. These sometimes compare well to impedance transitions in the logs, but they can be misleading due to the lack of low frequency information in the seismic trace.

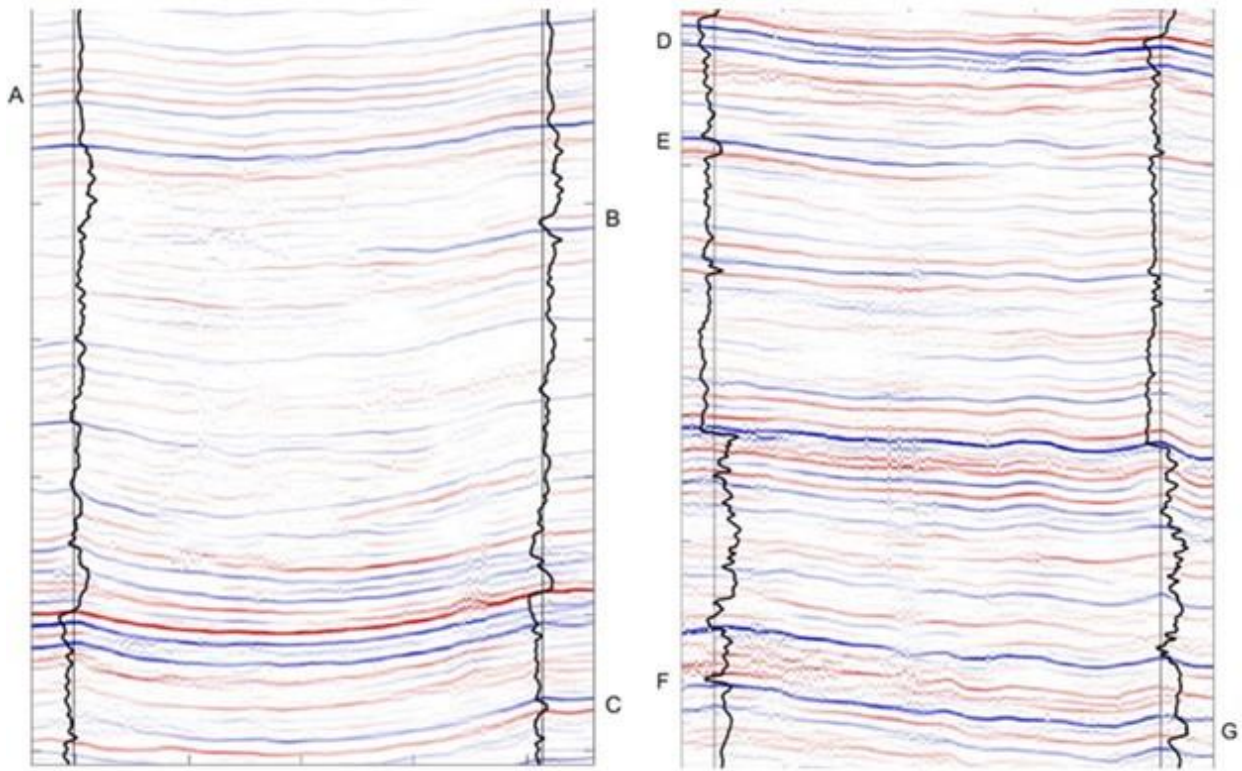

*Figure 7:* *Comparison of reflectivity inversions without a known or extracted wavelet or starting models to impedance logs (well log impedance increases from left to right). A - low impedance layer pinches out to the right.* ***B*** *- Layers pinch out to the left. C - Sharp top of high impedance layer becomes transitional to the left. D – Transition indicated by series of same sign reflection coefficients. E – high impedance layer pinches out to the right. F- Low impedance layer pinches out to the right. G – Layers diminish in contrast to the left. Red = negative(soft) reflection coefficients. Blue = positive (hard) reflection coefficients.*

The inverted bandlimited impedance (Figure 8) obtained by integrating the inverted reflectivity ties very well to thin layers but overshoots and undershoots at boundaries of thick layers due to missing low frequencies. The seismic bandlimited impedance correlates very well to the bandlimited impedance at wells obtained by removing the well log impedance low frequency trend (Figure 9). Adding this low frequency trend to the seismic bandlimited inversion at the well produces an absolute impedance inversion (Figure 9) which compares favorably to the well log.

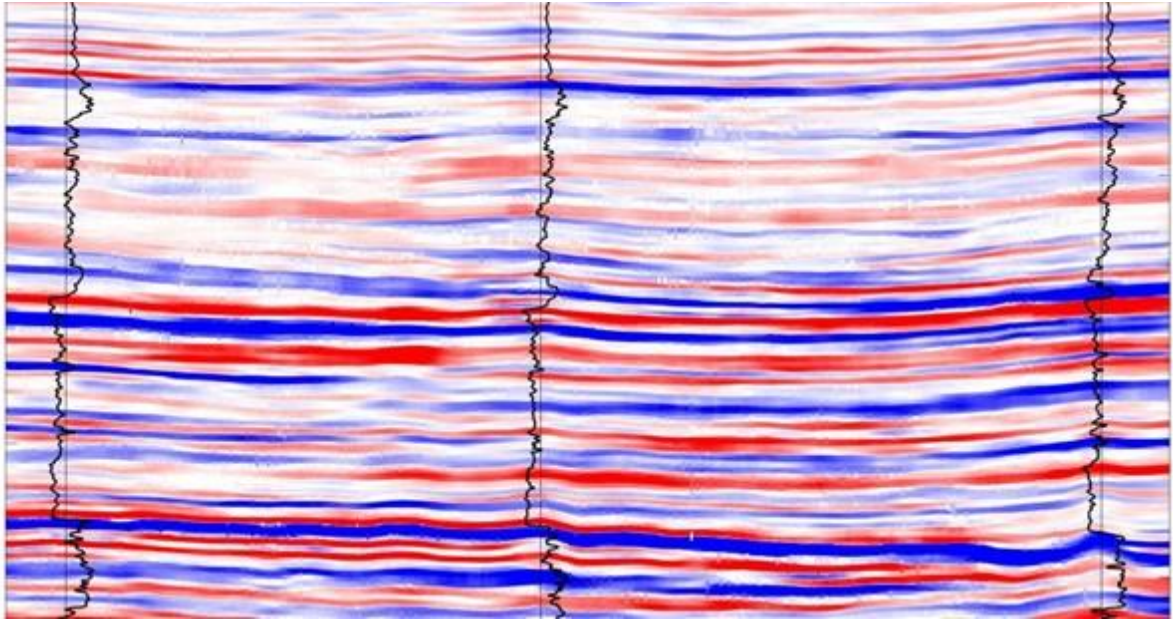

*Figure 8*: Reflectivity inversion of field seismic data with unknown wavelet at the well location. Inverted bandlimited impedance (red = soft, blue = hard) compared to impedance logs. Thin beds on seismic and well logs tie well but thick layers and large interfaces exhibit overshoots and undershoots due to missing low frequencies. These need to be added in to recover correct absolute impedance. Record time and spatial position is masked.

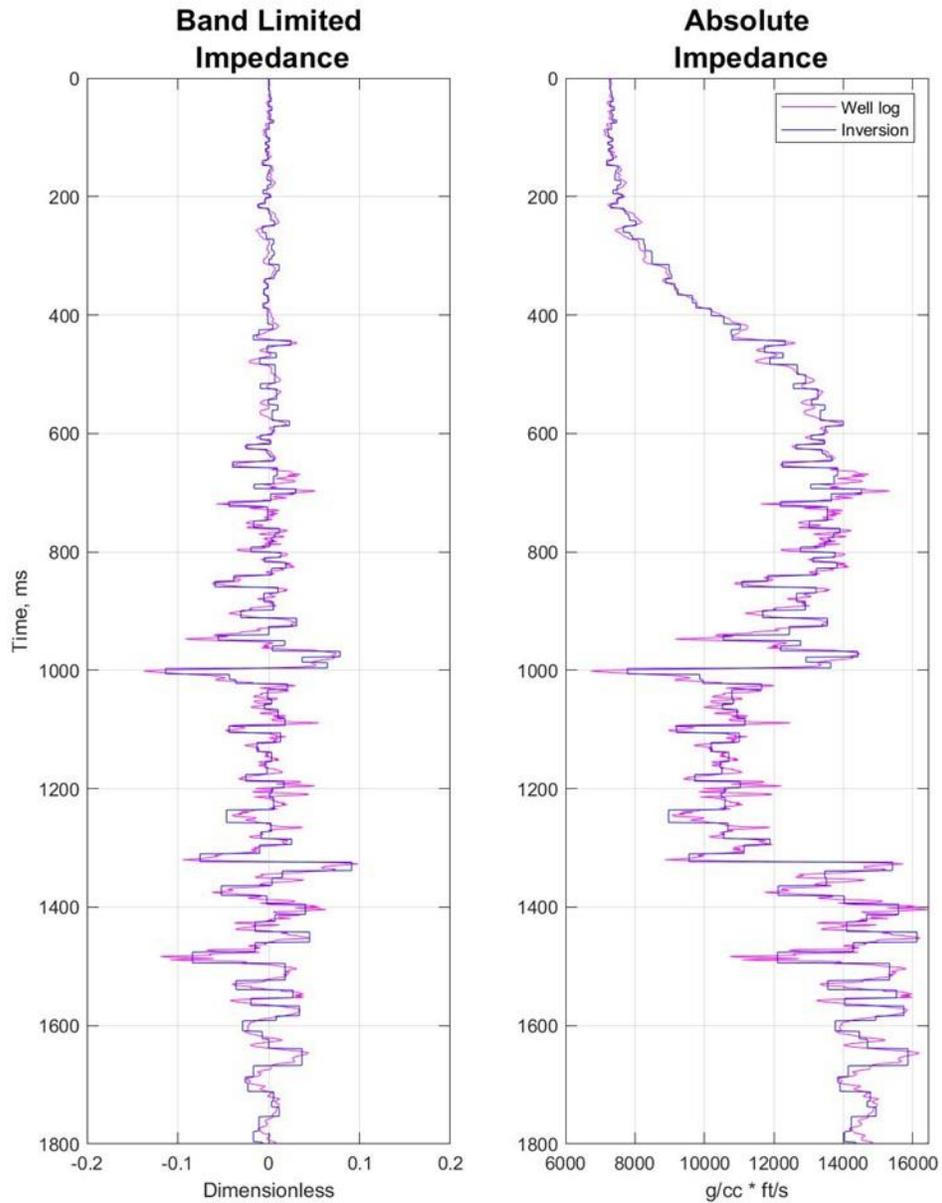

*Figure 9:* Bandlimited impedance (left track) and absolute impedance (right track) from well logs and seismic inversion without a known wavelet. Inversion bandlimited impedance is converted to absolute impedance by adding the low frequency impedance structure from the well log. Pink = impedance from well log, Blue = impedance from seismic inversion of field data without a known wavelet.

**Implied Wavelet Field**

A by-product of the inversion is the local time and space varying wavelet field implied by the inverted reflectivity. These implied wavelets are important for quality control of the process, as the wavelet phase variation is potentially the most problematic aspect of the procedure, especially when the true earth reflectivity is not sparse or when the local wavelet phase spectrum is complex. In this case, apparent non-physical variations in phase may occur and these should be discarded. Figure 10 shows the implied wavelets by inversion superimposed on the original seismic data shown in Figure 6. Although the data were carefully "zero-phased" in processing the wavelet phase can be seen to rotate from slightly negative to slightly positive with increasing record time (Figure 11). It is likely that most so-called zero-phase data has actual wavelet phase variations of this order or greater.

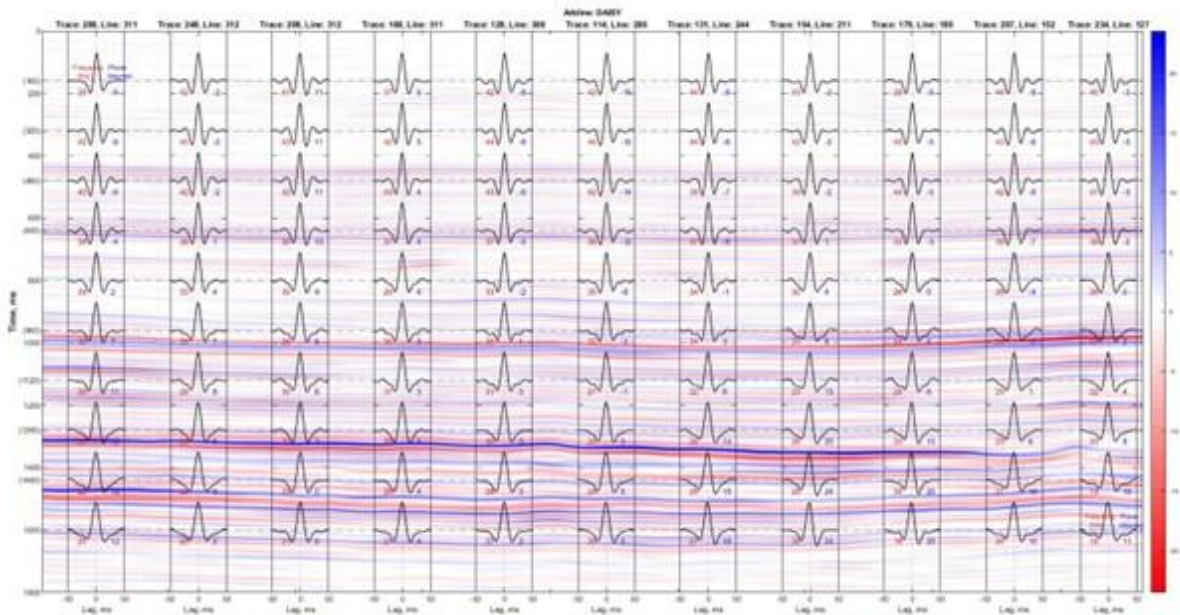

*Figure 10:* Implied wavelets output by the inversion superimposed on the original seismic data in the time and space location where they are locally derived. The blue numbers are wavelet phase, and the red numbers are wavelet dominant frequency. The loss of high frequencies and progressive phase rotation with record time is evident.

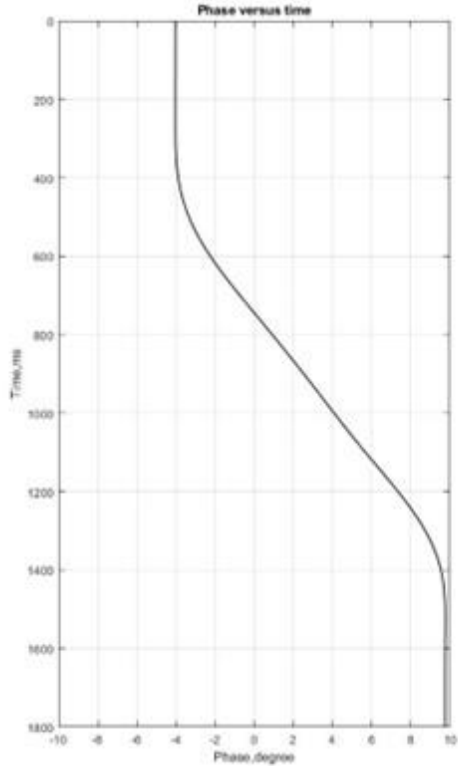

*Figure 11: Trend of average phase versus record time for the dataset depicted in Figure 10.*

**Inversion of Depth-Migrated Data**

As no local wavelet is required, the inversion can be run directly on depth-migrated data. Figure 12 shows an example in the vicinity of a well. Comparison of the inverted bandlimited impedance at the well to the well log bandlimited impedance shows a reasonable but not outstanding correlation. Nevertheless, comparison to the logs suggest that the reflectivity inversion has better apparent resolution than the original data. The inverted bandlimited impedance could be sufficient to establish misties between imaging and well log depths and may therefore aid in velocity model building. In depth migration, local asymmetrical stretching and squeezing over the length of the wavelet caused by rapid velocity variations can produce effective wavelet phase spectra that are not well described by a constant wavelet phase rotation. Evaluation and mitigation of this problem is the objective of ongoing research.

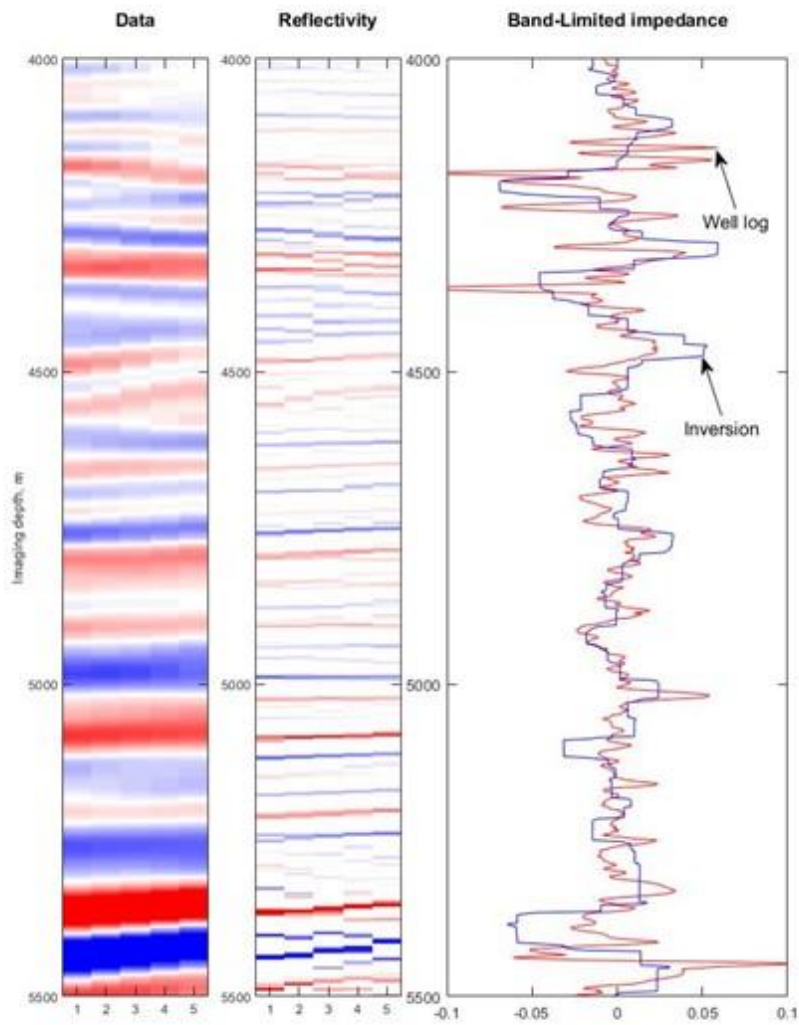

*Figure 12*: Wavelet-free inversion applied directly to depth migrated data at a well. The first track is the original seismic data, the middle track is the inverted reflectivity, and the right-most track is the inverted bandlimited impedance at the well compared to the well log bandlimited impedance.

**Conclusions and Discussion**

Reflectivity inversion of seismic data is possible without explicit *a priori* knowledge or extraction of the seismic wavelet. The process involves the direct sparse geophysical inversion of features of the Multiscale Fourier Transform. No training is required for application to a particular dataset. The first inversion pass determines the local amplitude spectrum of the reflectivity and thus implies the local seismic wavelet amplitude spectrum. If the true seismic wavelet were zero phase, the inversion at this point could be accepted. However, the locally inverted reflectivity phase spectrum will inherit the phase of the seismic wavelet. This is handled by reinverting with variable phase rotations of the implied zero phase wavelet obtained from the first inversion pass subject to physical constraints. The phase rotation that yields the sparsest reflectivity is accepted as the local implied wavelet phase, and the data are re-inverted using the time and space varying implied wavelet field. Display of this wavelet field is an important quality control step.

The inversion algorithm used here requires setting of a regularization parameter to control the tradeoff between sparsity and accuracy. In the algorithm we used here, there are no input parameters as we employ a hard-wired relation between the regularization parameter used and the information content of the seismic trace. Thus, lower frequency and narrower bandwidth data produce sparser solutions as can be seen in Figure 3. It is of course possible to override this automatic determination if, for example, greater resolution is desired and there is tolerance for some more inversion instability.

The ability to invert seismic data without prior knowledge of the seismic wavelet greatly enhances the inversion process. The need for synthetic ties and wavelet extractions prior to inversion is not only time consuming, and thus expensive, but the inherent inaccuracy in that process (because the wavelet is poorly known and varies with time, position, and offset) is largely swept under the rug in conventional practice. Problems with conventional methods of determining the seismic wavelet include (1) statistical wavelets are corrupted by the reflectivity spectrum, (2) well log derived wavelets are harmed by well log errors and imperfect time-depth functions, (3) the synthetic well tie process is circular because the wavelet is required to establish the synthetic tie needed to extract the wavelet, (4) there is a trade-off between seismic reflection arrival time and the determined wavelet phase, (5) using well log reflectivity to extract the wavelet usually results in different wavelets extracted at different wells and how those differences should be handled is often ambiguous, and (6) depth migrated data has wavelets that vary with velocity while technically incorrect cosmetic post-migration processes such as bandpass filtering in depth distort the wavelet, and (7) there is no objective way to verify the extracted wavelet between wells. As a result, constant wavelets are frequently used with the, perhaps naively optimistic, hope that spatial and temporal wavelet variations are adequately corrected in processing. The inversion method described here does not require the wavelet to be known in advance and thus is not perturbed by actual wavelet variations with time, spatial position, and offset.

The Multiscale Fourier Transform is an essential element of the described method. Without it, relative amplitudes between window locations and lengths would not be preserved. The role of the neural network training we applied was only to identify the essential wavelet independent features of the Multiscale Fourier Transform. It theoretically need never be applied again as the inversion is purely a geophysical procedure based on a forward model that matches the inverted earth model to the spectral features. It is possible that inversion performance could be improved for a particular dataset by retraining and optimizing the features to be used, but we do not expect this to be an essential step.

The weakest link in the wavelet-free inversion process is the determination of implied seismic wavelet phase which relies on the assumption that reflectivity is sparse – i.e., that the earth impedance structure is blocky. Deviation from this assumption can cause error in the phase determination as can actual wavelet phase that is not nearly constant with frequency over the band of the seismic data. Inspection of implied wavelets, as well as deglitching, smoothing, and interpolation of them is an important quality control step that can be followed by a final inversion using the spatially and temporally varying edited wavelet field. The wavelet field editing process can be aided by synthetics using the time-varying wavelets determined at any well location, however, no wavelet editing was performed in the examples shown.

As a sparse inversion, the reflectivity inversion described here does not require a starting model and is thus objective and independent of interpreter bias.  This is important because starting models are sometimes inaccurate and pull the inversion result towards the initial biased guess.  Interpolated well logs are biased towards lateral continuity and even low frequency variation between wells with correct check-shot calibration can be wrong if velocities vary laterally differently than a strict interpolation would suggest.  Further, interpolation between wells along horizons requires correct horizon picks which can be misinterpreted. In everyday practice, these are usually snapped to a peak, trough, or other part of the waveform, while actual horizon tops may fall anywhere on the waveform; this location varying with lateral changes in rock properties and other stratigraphic variations.  Correct horizon tops are better determined on the reflectivity inversion than on the original seismic data.

Even if one is committed to performing a conventional seismic inversion, the proposed method provides a rapid quick look diagnostic which can reveal critical issues, implied wavelets, and proper picking of formation tops.  It requires no extracted wavelet, no synthetic ties, no time-depth functions, no horizons, and no starting model.  It can be run directly on depth data and on pre-stack gathers.  The output reflectivity and bandlimited impedance allow for improved geological visualization and interpretation relative to the original seismic data. As merging with a low frequency earth impedance model is still required to get quantitative estimates of absolute impedance, attachment to elastic 3D full-waveform inversion (FWI) to perform direct waveform inversion (DWI) is an obvious next step.